\documentclass[sigconf]{acmart}
\renewcommand\footnotetextcopyrightpermission[1]{}
\usepackage{booktabs} 
\usepackage{tabularx}
\usepackage{multirow}
\usepackage{amsmath}
\usepackage{graphicx}
\usepackage{subcaption}
\usepackage[utf8]{inputenc}
\usepackage{algorithm}
\usepackage[noend]{algpseudocode}

\begin{document}
\title{An AI aid to the editors}
\subtitle{Exploring the possibility of an AI assisted article classification system}

\author{Tirthankar Ghosal, Rajeev Verma, Asif Ekbal, Sriparna Saha, Pushpak Bhattacharyya}
\email{(tirthankar.pcs16,rajeev.ee15,asif,sriparna,pb)@iitp.ac.in}

\begin{abstract}
This work is a preliminary exploratory study of how we could progress a step towards an AI assisted article classification system in academia. The proposed system aims to aid the journal editors in their decisions by pinpointing the potential weaknesses or strengths of a submitted manuscript.  From a large collection of articles and corresponding author-editor interactions we explore the possible reasons that lead to a paper being not forwarded for review. Our investigation reveals that in most cases either it is because the prospective manuscript is out of scope of the journal or the manuscript does not satisfy the minimum quality requirements to maintain the standard of the journal. We extract several features to quantify the quality of a paper and the degree of in-scope exploring keyword search, citation analysis, reputations of authors and affiliations, similarity with respect to accepted papers. With these features we train standard machine learning classifiers to develop a classification system. On a decent set of test data our approach yields promising results across 3 different journals. We believe that our approach is generic and could be adapted to other journals with appropriate adjustments.
\end{abstract}

%
%



\maketitle

\section{Introduction}
With the ever expanding volume of research articles and the number of submissions made, it is increasingly becoming difficult for the editors to keep up with the pace of latest research, manually go through each submission, respond to the author(s) or forward to the expert reviewers in a reasonable time frame. The editor is usually overwhelmed with this "burden of science". Here in this work we take up the problem and seek to investigate \textit{how could we use artificial intelligence to classify articles based on certain factors} so that the "burden of science" on the editors eases to some extent. Our intention is not to develop an automated system (which is by far not desirable) but to make best use of scholarly content and meta data available to develop an assistive support system for the editors in general. We begin our investigations on a set of accepted and not-forwarded papers along with their author-editor-reviewer interactions, made available to us by the reputed scholarly publishing house, Elsevier. In an usual setup, the editor is the person acting as the intermediary between the authors and the reviewers. Hence our obvious point of departure was the analysis of those editorial communications with the authors and the reviewers. We seek to discover features from not-forwarded (NFWD) articles and the corresponding editorial communications to identify the reasons for non-acceptance. 
Rigorous analysis of about 5500 NF data from 11 different Computer Science journals led us to believe that apart from some journal specific factors, there exists at least 5 generic features that leads to an article being withheld : 
\begin{enumerate}
  \item Appropriateness of the article to the journal being sent (Aim and \textbf{Scope}).
  \item \textbf{Quality}/Standard and impact of the article under review.
  \item Percentage overlap with existing articles (Plagiarism).
  \item Spelling, grammar and language of the article under review.
  \item Visually discriminative features of the article such as template mismatch (article not being prepared according to journal guidelines and formatting requirements), articles not having the standard components of a proper scientific communication\footnote{We had to literally come across submissions that were some fancies about time machine and doomsday theory}.
\end{enumerate}
While there are significant standard systems available for (3), (4) and (5), we felt that (1) and (2) are the ones that deserve special research attention. In the subsequent discussions we uncover a wide range of features ranging from author credibility to citations to content to arrive to a system that performs a binary classification task : FORWARDED (FWD) or NOT-FORWARDED (NFWD). However our consensus is that this piece of work is merely a direction towards an \textit{AI assisted article classification system}. 
We perform appropriate data and feature analysis to come up with important observations for the academia. The proposed system performs fairly well with features investigated for (1) and (2). However with added features encompassing (3), (4) and (5), the system could be more accurate to classify articles. 
Authors could also benefit from such a system as they could try in and check the potential of their manuscript.
\begin{figure}[ht]
\includegraphics[height=6cm, width=\linewidth]{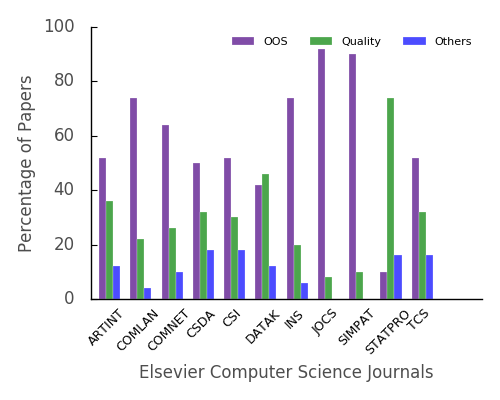}
\vspace{-6 ex}
\caption{Analysis of Not-Forwarded (NFWD) papers due to \textit{out-of-scope (OOS) and quality(Q)} for 11 different Elsevier Computer Science Journals. 500 NF papers for each journal were considered in the analysis}\label{fig:DR_stat}
\vspace{-3 ex}
\end{figure}
\section{Data}
\begin{figure*}[ht]
  \begin{subfigure}[b]{0.24\textwidth}
    \includegraphics[width=\textwidth]{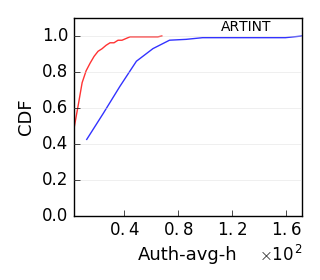}
    \caption{}
    \label{Auth-avg-h}
  \end{subfigure}
  \begin{subfigure}[b]{0.24\textwidth}
    \includegraphics[width=\textwidth]{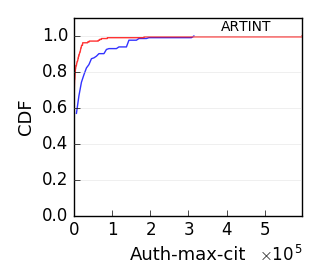}
    \caption{}
    \label{Auth-max-cit}
  \end{subfigure}
  \begin{subfigure}[b]{0.24\textwidth}
    \includegraphics[width=\textwidth]{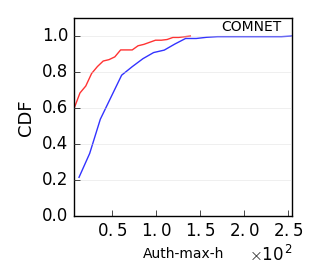}
    \caption{}
    \label{fig:2}
  \end{subfigure}
    \begin{subfigure}[b]{0.24\textwidth}
    \includegraphics[width=\textwidth]{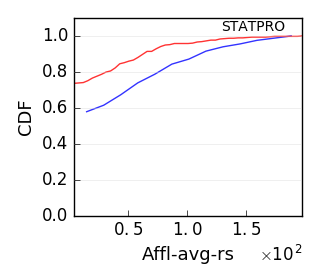}
    \caption{}
    \label{Affl-avg-rs}
  \end{subfigure}
  \begin{subfigure}[b]{0.24\textwidth}
    \includegraphics[width=\textwidth]{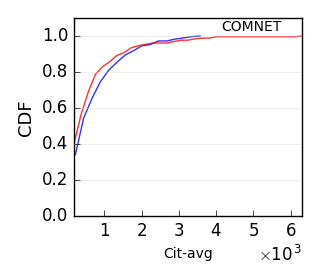}
    \caption{}
    \label{avgCit}
  \end{subfigure}
  \begin{subfigure}[b]{0.24\textwidth}
    \includegraphics[width=\textwidth]{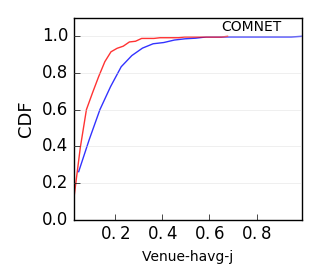}
    \caption{}
    \label{venuehavgj}
  \end{subfigure}
  \begin{subfigure}[b]{0.24\textwidth}
    \includegraphics[width=\textwidth]{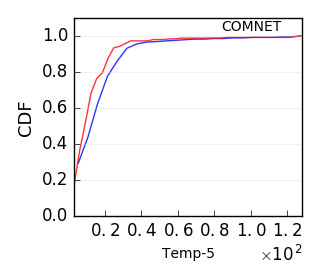}
    \caption{}
    \label{temp5}
  \end{subfigure}
   \begin{subfigure}[b]{0.24\textwidth}
    \includegraphics[width=\textwidth]{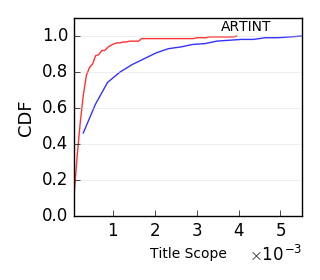}
    \caption{}
    \label{tit_sc}
  \end{subfigure}   
  \caption{Cumulative distribution function of various factors across \color{blue}{FWD} \color{black}and \color{red}{NFWD} \color{black}papers}
  \label{fig:1}
\end{figure*}
To address the task, we use a collection of 15,536 accepted papers from Elsevier Science Direct\footnote{https://www.sciencedirect.com/} pertaining to 3 different journals. Along with we consider and review a set of 5500 NF papers pertaining to 11 different Elsevier Computer Science journals \textit{viz.,} \textit{Artificial Intelligence} (ARTINT), \textit{Computer Networks} (COMNET), \textit{Theoretical Computer Science} (TCS), \textit{Statistics and Probability Letters} (STATPRO), \textit{Computational Statistics and Data Analysis} (CSDA), \textit{Computer Languages, Systems and Structures} (COMLAN), \textit{Computer Standards and Interfaces} (CSI), \textit{Data and Knowledge Engineering} (DATAK), \textit{Information Sciences} (INS), \textit{Journal of Computational Science} (JOCS), and \textit{Simulation Modelling Practice and Theory} (SIMPAT). We also went through about 7000 review reports (author-editor-reviewer interactions) of accepted, NF and declined-after-review papers consisting of more than 2.1 million lines of review data to investigate the generic causes of non-acceptance and then develop a good set of features. However for the purpose of cross-validation experiments, we use only a subset of these FWD and NF papers and report the results. We carry out our machine learning experiments on 3000 articles from ARTINT, STATPRO and COMNET journals.
\subsection{Data Preparation and Preprocessing}
The current work is too data intensive and we invest a huge amount of time in preparing the data, collecting meta data from various external sources. After preprocessing is done on the data we create several meta files for each journal consisting of the following information. (1) We extract Author and Affiliation scores from Google scholar, Scopus\footnote{https://www.elsevier.com/solutions/scopus} and Times Higher Education Ranking\footnote{https://www.timeshighereducation.com/world-university-rankings/2017/} data. We manually correct the noises (resulted due to parsing) and do author name disambiguation to get the actual scores from these external sources.
(2) Venue scores were extracted from Scientific Journal Rankings (SJR)\footnote{http://www.scimagojr.com} and CORE\footnote{http://portal.core.edu.au/conf-ranks/} rankings. Here too we did cleaning and removal of noises to tap and map actual entries from the external sources.
(3) Publication year was missing from many of the references. We had to fix those by actually searching in the web.
(4) Our pdf paper parsing aid, is excellent in GeneRation Of BIbliographic Data (GROBID)\footnote{https://github.com/kermitt2/grobid}. But due to inherent limitations of GROBID to parse the body of the scientific paper efficiently, we had to literally scan each paper with its generated xml to extract our content features (discussed in Table 1) correctly.
(5) Citation counts for reference entries we collect from Google Scholar\footnote{https://scholar.google.co.in/} and ArnetMiner\footnote{https://aminer.org/}.
\subsection{\textbf{Data Characterization}} 
(1) \textit{\textbf{Author reputation and affiliation}} If author credibility and affiliation reputation are high, most likely the manuscript undergoes less rigorous reviews as compared to others \cite{DBLP:conf/jcdl/SikdarMGM17}. With obvious exceptions acceptance rates of such manuscripts are generally high. We take the average of \textit{h}-index of authors in each FWD/NFWD article and plot their Cumulative Distribution Function (CDF) (Figure 2(a)). Even presence of a co-author having a very high credibility score (\textit{h}-index or citation count) resonates the behavior (Figure 2(b,c)). Authors from reputed institutions do have an edge over others. We take the research scores of affiliations from \textit{Times Higher Education Ranking} data and plot in Figure 2(d).\\
(2)\textit{\textbf{Quality of bibliography}} The quality of references in the bibliography section has a profound influence on the overall quality of the manuscript under review. Citation count of reference articles and their publishing venue are two important indicators of their quality. We plot the average citation count (Figure 2(e)) of references and average of \textit{h}-indices of journal venues (Figure 2(f)) for each paper  and study the qualitative impact of bibliography. Even recency of citations play a role. Figure 2(g) shows the distribution of citations that are published within 5 years from the date of submission of each article across FWD and NFWD papers. \\
(3) \textit{\textbf{Aptness of references}} We also observe that references hold strong \textit{domain} information. We seek to investigate the distribution of occurrence of reference titles across FWD and NFWD articles. The frequency of cited titles were always high when computed across all FWD articles. Whereas the frequency of cited titles across NFWD articles were low. Figure 2(h) supports the fact.
In ARTINT FWD, 75\% of the papers have average \textit{h}-index of the co-authors less than 38.00 while the same number for FWD is found to be 9.12 (Figure 2(a)). For STATPRO, 25\% of the FWD papers have average affiliation research score greater than 59.52 while the same statistics for NFWD stands at 15.30 (Figure 2(d)). Similar distinction we observe with other factors across FWD and NFWD articles for different journals.
\vspace{-2 ex}
\section{Motivation}
Inspired from the characterization of our data, our modeling of a desk-evaded forwarded paper is driven by three key observations :\\
  (1) \textbf{Addressing Scope} : \textit{If a paper belongs to a certain domain, then majority of its references would fall in that domain} 
  (2) \textbf{Addressing Quality} : \textit{The better is the bibliography section of a paper, higher is the prospect of a quality content inside the body of the paper.} (3) \textbf{Content} : \textit{Some content characteristics and location of impact/in-domain citations exerts a "local influence" on corresponding sections of a scientific paper that greatly determines its credibility as well as scope.}
By "better" we mean high impact as well as recent citations. Again citations could be "important" or "incidental" as per \cite{DBLP:conf/aaai/ValenzuelaHE15}. We devise a way to consider a subset of citations that are more contributing to the paper under review and not just have been cited incidentally (Section 5.4.3).
\vspace{-2 ex}
\section{Feature Engineering}
We summarize the intuitions behind considering these features in following sections. Each of the features are normalized with their respective maximum value.
\subsection{Scope Features}\textit{Suitability} of a submitted article to the "scope" of the target journal is perhaps the most deciding factor for which articles gets declined from editors desk. In spite of having merit many articles are simply not forwarded because they do not possess the content to cater the audience of the respective journal. Survey on a sample set of NFWD papers belonging to different Elsevier journals reveals this (Figure 1). We go by our intuition as discussed in Section 4 and derive features from the bibliography section as well as from the paper content.
\subsubsection*{\textbf{Dictionary Lists}} For each of the journals, we create distinct lists for:
(1) authors (2) author-listed keywords (3) paper titles appearing in the bibliography section (4) conferences appearing in the bibliography section (5) journals appearing in the bibliography section
and count their respective frequency of occurrences across all accepted (ACC) papers. For ARTINT we processed 3821 ACC articles, for STATPRO  7489 ACC articles and for COMNET 4226 ACC articles. We multiply frequency of each entry in (3), (4) and (5) with their respective \textit{\textbf{Citation Effect}}.
Citation Effect (CE) is the number of times a certain bibliographic item is cited within a paper. We introduce Citation Effect to measure the relative importance of a given bibliographic entry within the body of a paper. However the context of importance could be \textit{scope} or \textit{quality}. We observe that with this behavior, the \textit{domain-specific keywords, journals, conferences} tops the list which we deem as a vital indicator for \textit{scope} of a journal. CE for a \textit{core-domain} citation is always greater than an \textit{allied-domain} citation. With (1) we seek to investigate how many times a certain author has published in a particular journal. More an author publishes articles belonging to a certain domain greater is the chance that her prospective next article would belong to the same domain. 
\subsubsection{\textbf{Keyword Match}} For each journal, we create separate lists of author given keywords (in the Keywords section) from all the accepted articles of ARTINT, COMNET, STATPRO and record their frequencies of occurrences. Upon sorting we found that the representative terms for each of the journals appear at the top. We design this feature to emphasize the containment and relative importance of the keywords in the candidate article with respect to the Keyword Dictionary. The value for this feature for a candidate article $Y$ is thus calculated as :
\vspace{-1 ex}
\begin{displaymath}
KWScore_Y = \frac{|KW_Y \cap KW_D|}{|KW_Y|} \times  \sum_{i=1}^{|KW_Y \cap KW_D|}f(K_i)
\end{displaymath}
\begin{itemize}
    \item $KW_{Y}$: is the set of author defined keywords in the candidate article $Y$
    \item $KW_{D}$: is the set of keywords in the Keyword Dictionary $D$
    \item $f(K_i)$: is the frequency of keyword $K_i$ as listed in $D$
    \item $K_i \in \{KW_Y \cap KW_D\}$
\end{itemize}
\subsubsection{\textbf{Title Scope}} From the accepted articles of each journal, we create separate lists of all paper titles that have appeared in the reference sections. We also record their frequencies of references from within the body of the individual article and occurrence across all the accepted articles. Thus, the value for an article title ($T_i$) in the exhaustive list is calculated as :
\vspace{-3 ex}
\begin{equation}\label{eq : tit.}
  Value\quad V(T_i)=\sum_{j=1}^{n}f_j (T_i)
\end{equation}
where $f_j (T_i)$ corresponds to the number of times title $T_i$ has been cited within an article $j$ and $n$ is the total number of accepted articles for any journal. From the exhaustive list of paper titles, we calculate the Title Score ($T_Y$) of a candidate article $Y$ as :
\vspace{-2 ex}
\begin{displaymath}
 T_Y=\sum_{i=1}^{m}V(T_i)
\end{displaymath}
where $m$ is the total number of references in $Y$. $V(T_i)$ is calculated from Eq. \ref{eq : tit.}.
\subsubsection{\textbf{Conference Scope}}
From the accepted articles of each journal, we create lists of conferences in which articles referenced by the accepted papers of the corresponding journal are published. We also record the frequency of appearance of such conferences in the reference section of the accepted articles. Thus the value for a conference ($C_i$) in the exhaustive list is calculated as :
\vspace{-2 ex}
\begin{equation}\label{eq : conf.}
  Value\quad V(C_i)=\sum_{j=1}^{n}f_j (C_i)
\end{equation}
where $f_j (C_i)$ corresponds to the number of times conference $C_i$ appears in the reference section of an article $j$ and $n$ is the total number of accepted articles for any journal. Similarly from the exhaustive list of conferences, we calculate the Conference Score ($C_Y$) of a candidate article $Y$ as :
\begin{displaymath}
 C_Y=\sum_{i=1}^{m}V(C_i)
\end{displaymath}
where $m$ is the total number of conference references in $Y$. $V(C_i)$ is calculated using Equation \ref{eq : conf.}.
\subsubsection{\textbf{Journal Scope}} 
Likewise from the exhaustive list of journals we calculate the Journal Score ($J_Y$) of a candidate article $Y$ as :
\vspace{-2 ex}
\begin{displaymath}
 J_Y=\sum_{i=1}^{m}V(J_i)
\end{displaymath}
where $m$ is the total number of journal references in $Y$. 
\subsubsection{\textbf{Author Domain Publication Frequency (ADPF)}} From the accepted articles, we record the publication frequency of authors in those three journals separately. More an author publishes articles belonging to a certain domain greater is the chance that her prospective next article would belong to the same domain. For a candidate article, we take the summation of the publication frequency of each of the authors in the journal concerned from the author list. In case of a new author publishing in that particular field it is unlikely that all the other authors are from different field/domains. Summation helps in ignoring the effect of a new-author.  
\subsubsection{\textbf{Cluster Distance}}
We observe that the ACC articles of a certain journal could be grouped into clusters representing different sub-domains within the journal scope. 
Thus the distance of a given research article from the set of clusters formed on the ACC articles may contribute to determine its scope. Any outliers to such clusters may be considered as \textit{out-of-scope}. With this intuition we perform the steps in Algorithm \ref{alg:cluster}.
\begin{algorithm}
\caption{Distance from cluster boundary}\label{alg:cluster}
\begin{algorithmic}[1]
\State Use RAKE\cite{rose2010automatic} 
to automatically extract Keywords
from the Title, Abstract, Introduction and Conclusion sections of an article $Y$ belonging to journal $J$.
\State Use \textit{word2vec}\cite{mikolov2013distributed} to generate the vectors of the extracted keywords (top 30 
RAKE extracted keywords) from $Y$.
\State Calculate the document vector of $Y$ by concatenating all the keyword vectors from \textit{Step 2}.
\State Repeat \textit{Steps 1-3} for all the accepted articles of the journal $J$.
\State Use Word Mover's Distance (WMD)\cite{kusner2015word} as the distance metric between two document vectors and generate the similarity matrix.
\State Apply K-Medoids\cite{kaufman1990partitioning} on the similarity matrix from \textit{Step 4} to generate the clusters ($C_i$) [K is determined via Silhouette Index; user tune-able; can vary across journals]
\State Find the radius($r_i$) of a cluster $C_i$ as:
\begin{displaymath}
 r_i=median(distance(c_i , p_j))
\end{displaymath}
where $c_i$ is the centre of cluster $C_i$ and $p_j$ is any point within cluster $C_i$.
\State Find the document vector ($p_Y$) of a candidate article $Y$ using \textit{Steps 1-3}.
\State Distance of the candidate article $Y$ from the boundary of cluster $C_i$ is given as :
\vspace{-2 ex}
\begin{displaymath}
 D_i=distance(c_i , p_Y) - r_i
\end{displaymath}
\State Repeat \textit{Step 9}  for all the clusters ($C_i$) obtained from \textit{Step 6} to get :
\vspace{-2 ex}
\begin{displaymath}
 D_Y=minimum(D_i)
\end{displaymath}
\end{algorithmic}
\end{algorithm}
We take \textit{minimum} of the distances of the candidate article $Y$ from the cluster centers, in order to learn how close is $Y$ to any of the clusters so formed. 
\subsubsection*{\textbf{Computer Science Specific Word Embeddings}}\textit{One major contribution in executing this feature is the creation and usage of \textbf{word2vec}\cite{mikolov2013distributed} word vectors trained on the entire Computer Science journal articles of Elsevier (to preserve domain dependency)}. We processed 41737169 sentences from around 400K articles. The embedding dimension was set to 300. We chose lines of texts extracted from \textit{Title, Abstract, Introduction, Body, Conclusions} sections of ACC articles pertaining to all Computer Science journals of Elsevier. Certain preprocessing needs were : removal of special characters, headings, table and figure captions, etc. We considered only complete sentences.
\subsection*{Quality Features}
\subsection{Author Features}
These are the features that determine the credibility of the author(s). We design our features based on pure observation of data and try to minimize the bias by taking different facets of author credibility. The \textit{h}-index of an author indicates the quality and the impact of the work he/she has done. A scholar with an index of \textit{h} has supposedly published \textit{h} papers, each of which has been cited in other papers at least \textit{h} times. Again the total citations received by an author sometimes works as an indicator for his/her research impact. Since primary authors are the ones who do the actual experiments and lead the collaboration, we take their past academic impact (\textit{Auth-P-h, Auth-P-cit}) as features. Again many a times primary authors are students under some supervisors who are usually co-authors. Quite naturally student authors would suffer in \textit{h}-index and citation counts. This is compensated by taking the \textit{average} and \textit{maximum} of the citation count and \textit{h}-indices of all the authors (\textit{Auth-max-h, Auth-max-cit,Auth-avg-h, Auth-avg-cit}). Information for these features we gather from the academic search engine \textit{Google Scholar} and the academic dataset from ArnetMiner. We normalize the feature values by dividing with the corresponding maximum.
\subsection{Affiliation Features}
Although not desired, still we found that affiliation of the author(s) turns out to be a major indicator. However we believe that even a relatively new author from a non-premier institution produces a quality work. Several other factor(s) like content, bibliography would then contribute towards its quality.  On the contrary an established author from a premier institution could get his paper refuted if it fails to satisfy those quality metrics. The usage of author and affiliation features as a measure of scientific impact has already been studied in \cite{DBLP:conf/wsdm/DongJC15,DBLP:conf/jcdl/YanHTZL12,livne2013predicting}. In order to measure \textit{Research Score} of an affiliation, we consider  the data pertaining to 2017 global ranking of different institutes published by \textit{Times Higher Education World University Rankings}\footnote{https://www.timeshighereducation.com/} which is usually taken as a standard in this field(\textit{Affl-P-rs, Affl-avg-rs, Affl-max-rs}). We deem number of research documents produced or published by an institution as a qualitative measure of its research ambition and potential. We take the indexed count from \textit{Scopus} which is again a standard source in this field. Academicians associated with premier institutions usually exhibit research potential that cumulates to a measure of overall research prowess of the institution. Here also we consider measures for primary authors, average and maximum among the authors (\textit{Affl-P-doc, Affl-avg-doc, Affl-max-doc}). Intuitively we rely on this observation that a reputed co-author would never want to get associated with an inferior quality work and would allow submission only when the prospective manuscript adheres to certain quality benchmark.
\subsection{Bibliographic Features}Our intuition on working with the bibliographic section of a prospective manuscript followed from this observation : "\textit{A quality paper would always refer to quality papers}". 
\subsubsection{\textbf{Venue}} The most robust measure of quality is the venue in which the referred paper got published. Quality of a journal venue is measured in terms of \textit{h}-index and \textit{Scientific Journal Rankings}\footnote{http://www.scimagojr.com/} (SJR) score produced by SCImago Lab powered by \textit{Scopus}. These are well-accepted scores harboring the quality of a journal. We take the average of the \textit{h}-index and SJR values as features (\textit{Venue-havg-j,Venue-sjravg-j}). For conferences we refer to the CORE\footnote{http://portal.core.edu.au/conf-ranks/} rankings. The CORE Conference Ranking is an ongoing activity that provides assessments of major conferences in the computing disciplines. We specify numerical importance to the conferences based on their CORE grades of 2017 as A*$\rightarrow$6, A$\rightarrow$4, B$\rightarrow$3, D$\rightarrow$2. 
We deem that conferences not listed in CORE are likely to have lesser stringent criteria in terms of quality. So we leave them from our calculations. We take the average of the CORE rankings of the referred conferences as a feature (\textit{Venue-avg-m}).
\subsubsection{\textbf{Temporal Distance}} The Temporal distance features measure whether a prospective submission has cited recent works. They are strong indicators of whether the author(s) are aware of the current \textit{state-of-the-art}. We calculate the Temporal Distance (TD) as :
\begin{equation} \label{TD}
TD_{YR}=Y_{year}-R_{year}
\end{equation}
where $TD_{YR}$ is the temporal distance of candidate article $Y$ from referenced article $R$. $Y_{year}$, $R_{year}$ are the years of publication/submission of articles $Y$ and $R$ respectively. We thus take average and minimum of temporal distance of impact citations as features. (\textit{Temp-avg, Temp-min}). We also count the number of impact citations within a temporal distance 5 (\textit{Temp-5}). Very high \textit{Temp-avg, Temp-min} and low \textit{Temp-5} are not desirable.
\subsubsection{\textbf{Citation features}}Citation counts of referenced articles are also  good measures of the quality of references. But again (i) not all referenced articles are central to the core of the target paper. And (ii) the citation count measure is time-dependent. A very recent article with huge scientific impact will have low citation count as compared to one published 20 years back. In \cite{DBLP:conf/aaai/ValenzuelaHE15} authors have showed that many citations are just incidental and do not contribute directly to the theme of the paper. So a certain citation-rich reference, although incidental to the target paper may hugely effect this category of features, if used directly. To negate these constraints we develop a heuristic to identify impact citations i.e., citations which are central to the theme of the paper, upon the shoulders of which the target paper stands tall. \\
\textbf{Identifying Impact References}: To identify impact references for a paper $Y$ we : (1) determine the distribution of all citations in $Y$ in terms of their citation counts. (2) extract a set $S$ of those citations which are too distantly distributed via median absolute deviation (MAD). The remaining citations form a set $C$ (3) For each item $i$ in $S$ we do : if $TD_i$ $>$ 5 \&\& $CE_i$ $<$ 2, we remove $i$ from $S$ (4) The impact references we consider from $C$ $\cup$ $S$ \\
We also induce the \textit{Citation Effect} (CE), discussed earlier, into action and multiply the citation counts of the impact references with their corresponding CE to get the weighted citation count. For \textit{W-Imp-cit-avg} feature we thus take the average of the weighted citation count of impact references. Also we keep \textit{Cit-avg} as another feature. Presence of \textit{Uncited references} is a negative quality for a paper and should be penalized.
\subsubsection{\textbf{Content features}} We take simple counts of \textit{tables, figures and mathematical equations} present within a paper as features (\textit{Math-eq, Table-c, Fig-c}). These structures generally aid in scientific understanding and makes the paper appear intuitive and logical.
\section{Experiments and Results}
As our problem was formulated as a classification task, we use a series of standard classifiers ranging from Support Vector Machines (SVM) 
to Logistic Regressions (LR)
to Multi Layer Perceptrons (MLP) 
with default parameter settings from the popular machine learning toolkit Weka.
We found that Random Forests (RF) 
perform best with our feature set across the 3 journals we experiment with. The reason could be attributed to the inter dependence of the features among themselves and presence of many outliers in our dataset. Also the scale of attribute values in our feature set is not uniform and varies to a large extent (For example, citation counts of some papers and authors were too high w.r.t. others). Random Forest is found to perform well in such situations. We employ  features extracted from 500 FWD and 500 NFWD papers from each of the three journals (ARTINT, COMNET, STATPRO) and train our classifiers. We report the 10-\textit{fold} cross-validation results as in Table 4. 
\begin{table}[h]
\begin{center}
\begin{tabular}{|c|c|c|c|c|c|c|} 
\hline 
 \multicolumn{3}{|c|}{\bf{ARTINT}} & \multicolumn{2}{|c|}{\bf{STATPRO}}
  & \multicolumn{2}{|c|}{\bf{COMNET}}\\
\hline 
\bf Features & \bf IG &  \bf R & \bf IG & \bf R & \bf IG & \bf R \\
 \hline \it{Keyword Match} & 0.32  & 1 & 0.11 & 2 & 0.14 & 7 \\  
 \hline
 \it{Cluster Distance} & 0.12  & 10 & 0.30 & 1 & 0.15 &  6\\ 
 \hline
 \it{Title Scope} & 0.21  & 3 & 0.0 & 26 & 0.16 & 5 \\ 
 \hline
 \it{Conference Scope} & 0.28  & 2 & 0 & 31 & 0.13 & 8 \\ 
 \hline
 \it{Journal Scope} & 0.09  & 14 & 0.05 & 8 & 0.03 & 26 \\ 
 \hline
 \it{ADPF} & 0.08  & 21 & 0.7 & 3 & 0.04 & 23 \\ 
 \hline
 \it{Auth-P-h} & 0.07  & 23 & 0.02 & 17 & 0.07 & 18 \\ 
 \hline
 \it{Auth-avg-h} & 0.20  & 4 & 0.02 & 14 & 0.19 & 2 \\ 
 \hline
 \it{Auth-max-h} & 0.14  & 8 & 0.01 & 21 & 0.17 & 4 \\ 
 \hline
 \it{Auth-P-cit} & 0.09  & 17 & 0.02 & 19 & 0.07 & 17 \\ 
 \hline
 \it{Auth-avg-cit} & 0.09  & 19 & 0.01 & 20 & 0.18 & 3 \\ 
 \hline
 \it{Auth-max-cit} & 0.15  & 7 & 0.0 & 28 & 0.21 & 1 \\ 
 \hline
 \it{Affl-P-rs} & 0.06  & 28 & 0.03 & 11 & 0.03 & 24 \\ 
 \hline
 \it{Affl-avg-rs} & 0.07  & 27 & 0.03 & 9 & 0.04 & 21 \\ 
 \hline
 \it{Affl-P-doc} & 0.09  & 18 & 0.05 & 7 & 0.06 & 20 \\ 
 \hline
 \it{Affl-avg-doc} & 0.09  & 16 & 0.05 &  6 & 0.07 & 19 \\ 
 \hline
 \it{Affl-max-doc} & 0.13  & 9 & 0.09 & 4 & 0.07 & 16 \\ 
 \hline
 \it{Affl-max-rs} & 0.07  & 24 & 0.03 & 10 & 0.08 & 15 \\ 
 \hline
 \it{W-Imp-cit-avg} & 0.15  & 6 & 0.02 & 13 & 0.11 & 10 \\ 
 \hline
 \it{Cit-avg} & 0.09  & 20 & 0 & 29 & 0.09 & 13 \\ 
 \hline
 \it{Temp-avg} & 0.10  & 13 & 0.03 & 12 & 0.03 & 25 \\ 
 \hline
 \it{Temp-min} & 0.07  & 26 & 0.0 & 25 & 0.02 & 27 \\ 
 \hline
 \it{Temp-5} & 0.09  & 15 & 0.01 & 23 & 0.10 & 11 \\ 
 \hline
 \it{Venue-havg-j} & 0.07  & 22 & 0.02 & 16 & 0.12 & 9 \\ 
 \hline
 \it{Venue-sjravg-j} & 0.03  & 29 & 0.01 & 24 & 0.02 & 29 \\ 
 \hline
 \it{Venue-avg-m} & 0.16  & 5 & 0.02 & 15 & 0.02 & 30 \\ 
 \hline
 \it{Uncited-ref} & 0.03  & 30 & 0.01 & 22 & 0.10 & 12 \\ 
 \hline
 \it{Math-eq} & 0.07  & 25 & 0.06 & 5 & 0.02 & 28 \\ 
 \hline
 \it{Table} & 0.11  & 12 & 0.02 & 18 & 0.10 &  22 \\ 
 \hline
 \it{Figure} & 0.12  & 11 & 0 & 27 & 0.09 & 14 \\ 
 \hline
\end{tabular}
\end{center}
\caption{Information Gain values for different features across  3 journals, $IG\rightarrow Information Gain$, $R\rightarrow Rank$}
\end{table}
\begin{table*}[h]
\begin{center}
\begin{tabular}{ |c|c|c|c|c|c|c|c|c|c|c|c|c|c|c|c| } 
\hline 
 \multicolumn{6}{|c|}{\bf{ARTINT}} & \multicolumn{5}{|c|}{\bf{STATPRO}}
  & \multicolumn{5}{|c|}{\bf{COMNET}}\\
\hline 
\bf Methods & \bf P(F) &  \bf R(NF) & \bf P(F)& \bf R(NF) & \bf A & \bf P(F) &  \bf R(NF) & \bf P(F)& \bf R(NF) & \bf A & \bf P(F) &  \bf R(NF) & \bf P(F)& \bf R(NF) & \bf A\\
 \hline \bf{RF} & 0.931  & 0.969 & 0.928 & 0.849 & \textbf{92.99} & 0.828  & 0.858 & 0.846 & 0.814 & \textbf{83.64} & 0.939  & 0.926 & 0.932 & 0.935 & \textbf{93.57} \\
 \hline
 \bf{LR} & 0.899  & 0.899 & 0.788 & 0.788 & 86.30 & 0.828  & 0.822 & 0.816 & 0.823 & 82.21 & 0.879  & 0.853 & 0.848 & 0.877 & 86.39\\ 
 \hline
 \bf{MLP} & 0.906  & 0.935 & 0.854 & 0.797 & 89.07 & 0.784  & 0.725 & 0.734 & 0.791 & 75.77 & 0.848  & 0.853 & 0.868 & 0.838 & 82.99\\ 
 \hline
 \bf{SVM} & 0.883  & 0.813 & 0.664 & 0.774 & 80.06 & 0.758  & 0.804 & 0.782 & 0.733 & 76.89 & 0.812  & 0.825 & 0.836 & 0.818 & 81.36\\ 
 \hline
\end{tabular}
\end{center}
\caption{\label{webis-cpc} Classification results for various journals, $P\rightarrow Precision$, $R\rightarrow Recall$,  $A\rightarrow Accuracy$ (in \%), $F\rightarrow Forwarded$,  $NF\rightarrow Not-Forwarded$}
\end{table*}
\subsection{Feature Significance}
\subsubsection{\textbf{ARTINT}} As is evident from Table 3, \textit{Scope}  features including \textit{Keyword Match, Conference Scope, Title Scope} appeared at the top. Artificial Intelligence is a rapidly changing domain and the current progress is mostly reflected in conference papers. Hence most of the citations found in ARTINT accepted papers were from premier AI conferences which contributed in scope determination. Also misfit submissions were identified with the \textit{Keyword Match} feature. Quality of venue (\textit{Venue-avg-m}) and quality of impact citations (\textit{W-Imp-cit-avg}) found prominence in the ranking signifying distinction between FWD and NFWD papers. However affiliation and temporal features did not factored that much. Even some content features like \# \textit{of tables or figures} contributed to some extent. We could vouch on this as we see a number of declined submissions to \textit{Artificial Intelligence} journal did not follow the usual criteria and standards of a proper scientific communication. Author credibility features like \textit{Auth-avg-h, Auth-max-cit, Auth-max-h} appeared in the top 10 justifying the power law distribution of scientific impact in current literature i.e. the "\textit{rich gets richer}".
\subsubsection{\textbf{STATPRO}} The \textit{Statistics and Probability Letters} journal has a narrow scope as compared to ARTINT and caters to a specialist audience interested in statistics and probability. We find that \textit{Cluster Distance, Keyword Match} feature emerged  as the most contributing one which is in line with the fact that STATPRO does not accept submission which has less similarity with respect to its history articles i.e., \textit{out-of-scope} submissions. Also analysis of data reveals that there are a limited set of specialist authors who contribute to STATPRO. Hence the ADPF feature is found important. Same reason goes for their \textit{Affiliation} features. The audience and contributors to STATPRO are generally from premier institutions and the discipline is not so wide as compared to ARTINT. So we see submissions to STATPRO are from a closed set of distinct authors belonging to a set of institutions. One particular content feature \textit{Math-eq} performed well due to obvious reasons. \textit{Temporal} features exhibited less relevance. 
\subsubsection{\textbf{COMNET}} Even the Computer Networks journal has a limited scope as compared to ARTINT. We find the \textit{Author} features to have predominance over others, however \textit{Affiliation} features did not work well. This signifies that authors for COMNET belong to varied institutions and that high impact authors too belong to non-premier institutions. Authors from non-premier institutions have quality submissions in COMNET. As expected \textit{Scope} features performed well. The reason being the domain of Computer Networks is well-defined and stringent as compared to ARTINT. \textit{Venue} feature \textit{Venue-havg-j} found it in the top 10 signifying that quality of citation venues, specifically journals, too had an impact on the classification decision. Impact citations also ranked among the top 10 features.
\subsubsection{\textbf{Feature Ablation}}
To better understand the effect of different category of features on our classification task, we ablate each category of features one by one and see the resultant average $F_1$ score across FWD and NFWD instances. Figure 3 clearly shows ablating \textit{Scope} features costs the maximum with a drop in $F_1$ by a margin of 9\% for ARTINT, 8\% for COMNET and 14\% for STATPRO. \textit{Author} features are distinct for ARTINT and COMNET while \textit{Affiliation} features are distinct for STATPRO. \textit{Citation, Venue, Temporal} features had little impact in presence of other features whereas \textit{Content} features displayed promise for STATPRO.
\begin{figure*}[ht]
  \begin{subfigure}[b]{0.3\textwidth}
    \includegraphics[width=\textwidth]{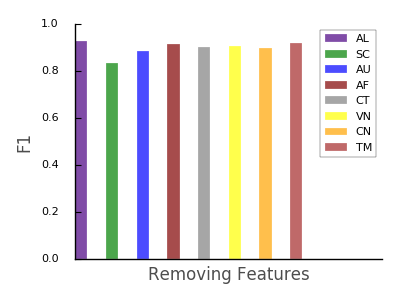}
    \caption{Feature ablation for ARTINT}
    \label{Auth-avg-h}
  \end{subfigure}
  \begin{subfigure}[b]{0.3\textwidth}
    \includegraphics[width=\textwidth]{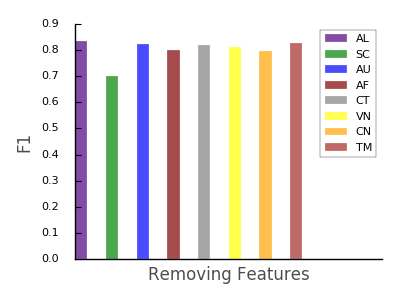}
    \caption{Feature ablation for STATPRO}
    \label{Auth-max-cit}
  \end{subfigure}
  \begin{subfigure}[b]{0.3\textwidth}
    \includegraphics[width=\textwidth]{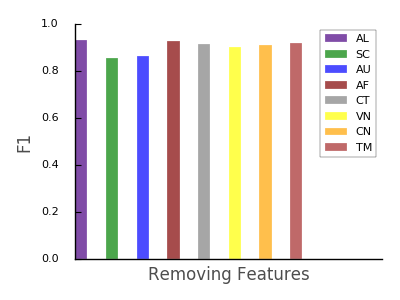}
    \caption{Feature ablation for COMNET}
    \label{fig:2}
  \end{subfigure}
  \caption{Feature contribution analysis. AL denotes training a Random Forest model with the full features set. The bars illustrate the average $F_1$ score after the corresponding category of feature is removed. SC $\rightarrow$ Scope, AU $\rightarrow$ Author, AF $\rightarrow$ Affiliation, CT $\rightarrow$ Citation features, VN $\rightarrow$ Venue, CN $\rightarrow$ Content, TM $\rightarrow$ Temporal : categories of features.}
  \label{fig:1}
\end{figure*}
\section{Summary}
According to all the analysis done, we summarize our findings, and provide the following intuitions to the academia : (1) A prospective article should always be judged against the topic relevance of the target journal before actually submitting. \textit{Authors should be aware of the domain and standards of publications hosted by the target journal}. (2) Presence of recent and quality \textit{Bibliographic items} and actually citing them in the manuscript inherently implies that the author must have done standard literature survey and is aware of the current \textit{state-of-the-art}. So \textit{a prospective author should read and refer to papers having good quality. Quality can be in terms of citation counts and author credibility for old papers. For recent papers one should definitely look into the venue credibility alongside author and affiliation}.(3) Although our approach considers \textit{Author} and \textit{Affiliation} past credibility as features, still we observe that only with those, his/her manuscript could not get past the editorial screening. The quality features should be strong enough as well. \\
\textbf{Case Study} : We did analysis of our approach on several test data. One such was for Paper Id : S0167715211003269\footnote{https://www.sciencedirect.com/science/article/pii/S0167715211003269\#!}. The author had a relatively low \textit{h}-index (1 to be precise) and do not belong to an elite institution as deemed by major institution ranking houses. Still due to sheer quality of contributions, citation-rich references and content factors, our classifier correctly predicted its accepted status. We also observe cases where a manuscript from a touted premier institution was not forwarded because of lack of quality of content and citations.
We restrain to report those due to confidentiality clauses.
\section{Conclusions}
Here in this work we deal with an interesting problem of bringing AI more close to the \textit{scientific editorial system}. 
The performance of our approach across journals of different domains bears testimony that our features are generic in nature as well as complementary to each other. There are many avenues where we could improve upon like mining information from content to decide upon the relevance and novelty of scientific claims, argumentation mining, 
more scientometric and qualitative analysis of bibliographic features, etc. Provided with enough data it would be interesting to see how \textit{Deep Learning} techniques would behave with this task. 

\begin{acks}
We are thankful to Elsevier for providing us the required data to conduct our experiments. These experiments are conducted in a usual academic lab setup with the sole intention to understand the nature of scholarly articles better, how AI could assist in taking better editorial decisions and overall how the scientific community could benefit from these studies. 

\end{acks}

\bibliographystyle{ACM-Reference-Format}
\bibliography{sample-bibliography} 

\end{document}